\documentclass[12pt]{article}
\usepackage{epsfig}
\usepackage{a4p}

\begin{document}
\def\be{\begin{equation} }
\def\ee{\end{equation} }
\def\ba{\begin{eqnarray} }
\def\ea{\end{eqnarray} }
\def\ban{\begin{eqnarray*} }
\def\ean{\end{eqnarray*} }
\def\epem{\mbox{e}^+\mbox{e}^-}
\def\eegg{\epem\to\gamma\gamma}
\def\eeggg{\epem\to\gamma\gamma(\gamma)}
\def\ggg{\gamma\gamma(\gamma)}
\def\ct{\cos{\theta}}
\def\cte{\cos{\theta^{\ast}}}
\def\pl{p_{\rm l}}
\def\pt{p_{\rm t}}
\def\g{\gamma}
\def\B{{\mathcal{B}}}
\def\O{{\mathcal{O}}}
\def\R{{\mathcal{R}}}
\def\E{{\mathcal{E}}}
\def\Lpm{\Lambda_{\pm}}
\def\xmc{\left(\frac{d\sigma}{d\Omega}\right)_{\rm MC}}
\def\xb{\left(\frac{d\sigma}{d\Omega}\right)_{\rm Born}}
\def\xl{\left(\frac{d\sigma}{d\Omega}\right)_{\Lambda_{\pm}}}
\def\xq{\left(\frac{d\sigma}{d\Omega}\right)_{\rm \Lambda '}}
\def\xe{\left(\frac{d\sigma}{d\Omega}\right)_{\rm e^{\ast}}}
\def\xg{\left(\frac{d\sigma}{d\Omega}\right)_{\rm G_{\pm}}}
\def\xsn{\frac{d\sigma}{d\Omega}}
\def\ca{$I$}
\def\cb{$I\!I$}
\def\cc{$I\!I\!I$}
\def\dd{$I\!V$}
\def\acol{\xi_{\rm acol}}
\def\acop{\xi_{\rm aplan}}
\def\ng{N_{\gamma}}

\begin{titlepage}
\begin{center}{\Large   EUROPEAN LABORATORY FOR PARTICLE PHYSICS
}\end{center}\bigskip
\begin{flushright}
       CERN-EP/99-088 \\
       1 July 1999
%       \today    \\
%       Journal Version, 5 August 1998
\end{flushright}
\bigskip\bigskip
\begin{center}{\LARGE\bf  
% titel 
\boldmath
Multi-photon production in $\epem$ collisions \\[1ex]
at $\sqrt{s} =$ 189 GeV 
\unboldmath
}\end{center}\bigskip\bigskip
\begin{center}{{\LARGE The OPAL Collaboration}

% Author: K. Sachs

% Editorial Board: K. Desch, J. Homer, G. Long, M. Oreglia 

}\end{center}\bigskip\bigskip

\begin{abstract}
The process $\eegg(\g)$ is studied using data recorded with the OPAL detector 
at LEP. The data sample taken at a centre-of-mass energy of 189 GeV corresponds 
to a total integrated luminosity of 178 pb$^{-1}$. 
The measured cross-section agrees well with the expectation from QED.  
A fit to the angular distribution is used to obtain improved limits at 
95\%\ CL on the QED cut-off parameters: $\Lambda_+ > $ 304 GeV and 
$\Lambda_- > $ 295 GeV
as well as a mass limit for an excited electron, $M_{\rm e^{\ast}} >$ 306 GeV
assuming equal $\rm e^{\ast}e\gamma$ and  $\rm ee\gamma$ couplings.
Graviton exchange in the context of theories with higher dimensions is excluded 
for scales $G_+ <$ 660 GeV and $G_- <$ 634 GeV.
No evidence for resonance production is found in the invariant mass 
spectrum of photon pairs. Limits are obtained for the cross-section times 
branching ratio for a resonance decaying into two photons and produced in
association with another photon.
\end{abstract}
\vspace*{1cm}

\begin{center}

{\large
(To be submitted to Phys. Lett.)
%        \\[2ex] {\Large Final Draft }
}

% Send comments to Kirsten.Sachs@cern.ch

% until June 26, 10:00, please

\end{center}
\end{titlepage}

\begin{center}{\Large        The OPAL Collaboration
}\end{center}\bigskip
\begin{center}{
%begin authorlist PLEASE DO NOT DELETE THIS COMMENT
G.\thinspace Abbiendi$^{  2}$,
K.\thinspace Ackerstaff$^{  8}$,
G.\thinspace Alexander$^{ 23}$,
J.\thinspace Allison$^{ 16}$,
K.J.\thinspace Anderson$^{  9}$,
S.\thinspace Anderson$^{ 12}$,
S.\thinspace Arcelli$^{ 17}$,
S.\thinspace Asai$^{ 24}$,
S.F.\thinspace Ashby$^{  1}$,
D.\thinspace Axen$^{ 29}$
% ...
%end authorlist PLEASE DO NOT DELETE THIS COMMENT
}\end{center}\bigskip
\bigskip
%begin institutes
$^{  1}$School of Physics and Astronomy, University of Birmingham,
Birmingham B15 2TT, UK
\newline
$^{  2}$Dipartimento di Fisica dell' Universit\`a di Bologna and INFN,
I-40126 Bologna, Italy
\newline
% ...
$^{  8}$CERN, European Organisation for Particle Physics,
CH-1211 Geneva 23, Switzerland
\newline
$^{  9}$Enrico Fermi Institute and Department of Physics,
University of Chicago, Chicago IL 60637, USA
\newline
% ...
$^{ 12}$Indiana University, Department of Physics,
Swain Hall West 117, Bloomington IN 47405, USA
\newline
% ...
$^{ 16}$Department of Physics, Schuster Laboratory, The University,
Manchester M13 9PL, UK
\newline
$^{ 17}$Department of Physics, University of Maryland,
College Park, MD 20742, USA
\newline
% ...
$^{ 23}$Department of Physics and Astronomy, Tel Aviv University,
Tel Aviv 69978, Israel
\newline
$^{ 24}$International Centre for Elementary Particle Physics and
Department of Physics, University of Tokyo, Tokyo 113-0033, and
Kobe University, Kobe 657-8501, Japan
\newline
% ...
$^{ 29}$University of British Columbia, Department of Physics,
Vancouver BC V6T 1Z1, Canada
\newline
% ...
 Debrecen, Hungary.
\newline

\section{Introduction}

This paper reports a study of the process $\eeggg$ using data 
recorded with the OPAL detector at LEP at an average centre-of-mass energy 
of 188.63 $\pm$ 0.04 GeV with an integrated luminosity of 178.3 $\pm$ 0.4 
pb$^{-1}$. This is one of the few processes dominated by QED even at high LEP 
energies. Since the differential cross-section is well known from QED
\cite{drell,mcgen}, any deviation from this expectation hints at
non-standard physics processes contributing to these photonic final states.
Any non-QED effects within the general framework of effective Lagrangian 
theory are expected to increase with centre-of-mass energy \cite{eboli}. 
A comparison of the measured photon angular distribution with the QED 
expectation can be used to place limits on the QED cut-off parameters 
$\Lambda_{\pm}$, contact interactions ($\epem\g\g$) and non-standard 
$\epem\g$-couplings as described in Section 3. 
The possible existence of an excited electron $\rm{e}^*$ which would 
change the angular distribution~\cite{litke}, is investigated
as well as the possibility of graviton exchange in the context of theories
with higher dimensions \cite{graviton}. In addition, a 
search is made for the production of a resonance X via 
\mbox{$\epem\to$ X$\gamma$} followed by the decay \mbox{X $\to \g\g$}, 
using the invariant mass spectrum of photon pairs in three-photon final states. 
The process $\eeggg$ has been analysed previously at lower energies 
\cite{ich172,ich183,l3,aleph,delphi}. 
The selection presented in this paper
is almost unchanged with respect to
the previous OPAL analysis at $\sqrt{s}$ = 183 GeV \cite{ich183}. 
% The following Section contains a brief description of the OPAL detector
% and of the Monte Carlo simulated event samples.
% Section 3 describes the QED differential cross-sections for $\eeggg$, as well
% as those from several models describing extensions to QED.
% In Sections 4 and 5 the analysis is described in detail. The results are
% presented in Section~6. 

\section{The OPAL detector and Monte Carlo samples}

A detailed description of the OPAL detector can be found in Ref. \cite{det}.
The polar angle $\theta$ is measured with respect to the electron beam
direction and $\phi$ is the azimuthal angle.
For this analysis the most important detector component is the 
electromagnetic calorimeter (ECAL) which is divided into two parts, the barrel 
and the endcaps. The barrel covers polar angles with $|\ct|<0.82$ and 
consists of 9440 lead-glass blocks in a quasi-pointing geometry. The two 
endcaps cover the polar angle range $0.81<|\ct |<0.98$ and each consists of 
1132 blocks. For beam-energy photons, the spatial resolution is about 11~mm, 
corresponding to 0.2$^{\circ}$ in $\theta$, and the energy resolution is 
about 2\%\ in the barrel and \mbox{3 -- 5\% } in the endcaps, depending on 
the polar angle. 
The ECAL surrounds the tracking chambers. Hit information from the central
jet chamber CJ and the central vertex drift chamber CV is used to reject 
events which are 
consistent with having charged particles coming from the interaction point. 
Outside the ECAL is the hadronic calorimeter HCAL which is incorporated 
into the magnet yoke, and beyond that are the muon chambers. Both the HCAL and 
the muon chambers are used to reject cosmic ray events.

Various Monte Carlo samples are used to study the selection efficiency and
expected background contributions.
For the signal process $\eeggg$ the RADCOR \cite{mcgen} generator is used
while FGAM \cite{fgam} is used for $\eegg\gamma\gamma$. FGAM does not take
into account the electron mass leading to a divergent cross-section at small 
angles and hence can not be used if at least one photon is along the beam axis.
The Bhabha process is simulated using BHWIDE \cite{mcbh} and TEEGG \cite{mcte}.
The process $\epem\to\overline{\nu}\nu\g(\g)$ is
%, $\mu^+\mu^-$ and $\tau^+\tau^-$ are 
simulated using
KORALZ \cite{mctt}. % PYTHIA \cite{mcmh} is used for hadronic events. 
All samples were processed through the OPAL detector simulation program 
\cite{mcdet} and reconstructed in the same way as the data.

\section{Cross-section for the process \boldmath $\eegg$ \unboldmath}

The Born-level differential cross-section for the process $\eegg$
in the relativistic limit of lowest order QED is given by \cite{bg}
\be
\xb = \frac{\alpha^2}{s}\;\frac{1+\cos^2{\theta} }{1-\cos^2{\theta} } \; ,
\label{born}
\ee
where $s$ denotes the square of the centre-of-mass energy, $\alpha$ is the
fine-structure constant and $\theta$ is the polar angle of one 
photon. Since the two photons are identical particles, the event angle is 
defined by convention such that $\ct$ is positive. 

In a multi-photon event, it is important to choose an appropriate definition of
the event angle. The event angle is not uniquely defined since the two
highest-energy photons in general are not exactly back-to-back.
The event angle $\cte$ used in this paper is defined as
\be 
\cos{\theta^{\ast}} = \left|\sin{\frac{\theta_1 - \theta_2}{2}}\right|
     \; {\Bigg /} \; \left( {\sin{\frac{\theta_1 + \theta_2}{2}}}\right) \; ,
\label{ctstar} \ee
where $\theta_1$ and $\theta_2$ are the polar angles of the two highest-energy
photons. This definition was chosen such that the deviations in the angular 
distribution with respect to the Born level are small, in this case between 
3 and 6\%\ for $\cte < 0.99$, as was shown in Ref. \cite{ich172}. For two-photon 
final states, $\cte$ is identical to $|\ct |$ and
for three-photon events in which the third photon is along the beam direction,
$\theta^{\ast}$ is equivalent to the scattering angle in the centre-of-mass 
system of the two observed photons. Since the angular definition is based on 
the two highest-energy photons, events with one of these escaping 
detection along the beam axis are rejected from the analysis.
The observed angular distribution is corrected for these higher-order effects
using ${\cal O}(\alpha^3)$ Monte Carlo studies in order that it can be directly 
compared to the following model expectations given in lowest order.

In Ref. \cite{drell}, possible deviations from the QED cross-section for Bhabha 
and M{\o}ller scattering are pa\-ra\-me\-tri\-sed in terms of cut-off parameters
$\Lambda_\pm$. 
These parameters correspond to a short-range exponential term added 
to the Coulomb potential. This ansatz leads to modifications of the 
photon angular distribution of the form 
\be
\xl   =  \xb \pm \frac{\alpha^2 s}{2\Lambda_\pm^4}(1+\cos^2{\theta}) \; .
\label{lambda} 
\ee
Alternatively, in terms of effective Lagrangian theory, a gauge-invariant 
operator may be added to 
QED. Depending on the dimension of the operator, different deviations from QED 
can be formulated \cite{eboli}. 
Contact interactions ($\g\g\epem$) or non-standard $\g\epem$ couplings 
described by operators of dimension 6, 7 or 8 lead to angular distributions 
with different mass scales $\Lambda$. In most cases these deviations are
functionally similar \cite{ich172}. Therefore, only the cross-section predicted 
by a \mbox{dimension-7} Lagrangian, given by
\ba
\xq & = & \xb + \frac{s^2}{32\pi}\frac{1}{\Lambda'^6} \; , 
\label{qed7} 
\ea
is studied here. Limits on mass scales for Lagrangians of dimension 6 and 8 
can easily be derived from limits on $\Lpm$ and $\Lambda '$ \cite{ich172}. 
 
The existence of an excited electron ${\rm e}^*$ with an
${\rm e}^*{\rm e}\g$ coupling
would contribute to the photon production process via $t$-channel exchange.
The resulting deviation from $\xb$ depends on the ${\rm e^*}$ mass $M_{\rm e^*}$ 
and the coupling constant $\kappa$ of the $\mathrm{e^* e\g}$ vertex relative 
to the $\mathrm{ee\g}$ vertex \cite{litke}:
\be \label{estar}
\left(\frac{d\sigma}{d\Omega}\right)_{\rm e^*} = 
\xb + f(M_{\rm e^{\ast}},\kappa ,s,\ct ).
\ee
The function $f$ is explicitly given in Ref. \cite{ich172}. 
In the limit $ M_{\rm e^*} \gg \sqrt{s}$, $\xe$ approaches $\xl$ with the
mass related to the cut-off para\-meter by 
$ M_{\rm e^*} = \sqrt{\kappa}\;\Lambda_+$.

Recent theories \cite{graviton} have pointed out that the graviton G might propagate
in a higher-dimensional space where additional dimensions are compactified
while other Standard Model particles are confined to the usual 3+1 
space-time dimensions. This would allow the large
Planck scale $M_{\rm Pl}$ with an effective scale $M_{\rm D}$ of order 
of the electroweak scale (${\cal O}(10^{2-3}$GeV))
 via $M_{\rm Pl}^2 = R^n \! M_{\rm D}^{n+2}$,
where $R$ is the compactification radius for $n$ higher dimensions. 
% The momentum component of the graviton in the extra dimensions appears as
% an effective mass in the usual four dimensional space-time. 
Although the contribution from
$\epem\to \rm G^{\ast} \to \gamma\gamma$ for a single mode is small compared to
the Standard Model contribution, the very large number of possible excitation
modes in the extra dimensions might lead to a measurable effect on 
the resulting differential cross-section, given by \cite{giudice}
\be \label{graviton}
\xg = \xb \pm \frac{\alpha s}{\pi^2} \; G^{-4}_{\pm}\;(1+\cos^2{\theta})
    + \frac{s^3}{64 \pi^2} \; G^{-8}_{\pm} \;(1-\cos^4{\theta}) \; ,
\ee
where $G_{\pm}$ are scale parameters of order of the 
effective scale $M_{\rm D}$.

\section{Event selection}

\subsection*{A) Preselection}
% {\bf A) Preselection:}
The preselection has only small changes compared to our previous analysis 
\cite{ich183}. Events are selected if they have two or more 
ECAL clusters within the polar-angle range $|\ct| < 0.97$ each with more 
than 1 GeV of deposited energy uncorrected for possible energy loss in the
material before the ECAL. Other energies used in the analysis are corrected 
for these losses. A cluster must consist of at least two lead-glass blocks.
This preselection is tightened by requiring a total energy in the ECAL  
of at least 10\% of the centre-of-mass energy. 

% {\bf B) Rejection of non-physics backgrounds:}
\subsection*{B) Rejection of non-physics backgrounds}
A cosmic-ray particle can pass through the hadronic and electromagnetic
calorimeters without producing a reconstructed track in the
central tracking chambers. The requirements to reject these events
are the same as in our previous analysis. 
% These particles do not cross the beam axis.
% Since the HCAL and ECAL have different radii, the resulting hits in the 
% two detectors occur separated in azimuth. To remove this background, 
We reject 
events with  3 or more track segments found in the muon chambers.
In the case of fewer than three such track segments, the
event rejection depends on the highest-energy HCAL cluster of the event.
Events are rejected if this HCAL cluster is separated from each of the photon
candidates by more than 10$^{\circ}$ in azimuth and has at least 1 GeV of
deposited energy in the case of one or two muon-track segments, or 
at least 15\%\ of the observed ECAL energy if no muon-track segments are found.

Another type of background is consistent with well localised electronic noise 
in the ECAL. The cuts used to reject these events are
slightly relaxed compared to our previous analysis to account for the
larger cluster size due to the higher centre-of-mass energy.
An event is rejected if one of these
accumulations consists of more than 14 ECAL clusters or has an extent of
more than 0.5 rad in $\theta$ or 0.5 rad/$\sin{\theta}$ in $\phi$.

% {\bf C) Kinematic requirements:}
\subsection*{C) Kinematic requirements}
The selection is based primarily on the requirement of small missing
energy and small missing transverse momentum and is unchanged with respect to
our previous analysis. Non-physics background and
events containing invisible particles or having only a small energy deposit in
the ECAL are rejected by this selection.
In particular, events from the Standard Model process 
$\epem\to\bar{\nu}\nu\gamma\gamma$ fail this selection.
Better resolution on quantities such as the total longitudinal or transverse
momentum of an event is obtained by using cluster angles and three-body
kinematics where possible, rather than using measured energies. Therefore the
event sample is divided into four classes, \ca -- \dd . The classes are
distinguished by the number of photon candidates $N_{\gamma}$, the 
acollinearity angle $\acol$, and the aplanarity angle $\acop$:
\ba 
\acol & = & 180^{\circ} - \alpha_{12}  \\
\acop & = & 360^{\circ} - (\alpha_{12} + \alpha_{13} + \alpha_{23}),
\ea
where $\alpha_{ij}$ is the angle between clusters $i$ and $j$ and 
the indices are ordered by decreasing cluster energy. 

\begin{table}[b]
\renewcommand{\arraystretch}{1.2}
\setlength{\tabcolsep}{1mm} 
\begin{center}
\begin{tabular}{|l|rcl|rcl|rcl|rcl|} \hline
Event class & \multicolumn{3}{c|}{\ca }&\multicolumn{3}{c|}{\cb }&
   \multicolumn{3}{c|}{\cc }&\multicolumn{3}{c|}{\dd}\\\hline
${\mbox{Number of}\atop\mbox{photon candidates}}$ \rule{0mm}{3.5ex}&
   $N_{\gamma}$ & $\geq$ & 2 & $N_{\gamma}$ & = & 2 & 
   $N_{\gamma}$ & = & 3 & $N_{\gamma}$   & $\geq$ & 3 \\
Acollinearity & 
   $\acol$& $<$ &$10^{\circ}$& $\acol$& $>$ &$10^{\circ}$ &
   $\acol$& $>$ &$10^{\circ}$& $\acol$& $>$ &$10^{\circ}$ \\
Aplanarity &
   & & & & & & $\acop$&$<$&$0.1^{\circ}$&$\acop $&$>$ &$0.1^{\circ}$ 
\\\hline
%   \\[-1.5ex]
% \multicolumn{13}{|c|}{~\dotfill ~}\\
Energy sum & 
   $E_S^I$& $>$ &$0.6 \sqrt{s}$ & $E_S^{I\!I}$& $>$ &$0.6 \sqrt{s}$ &
   $E_S^{I\!I\!I}$& $>$ &$0.6 \sqrt{s}$ & $E_S^{I\!I\!I}$& $>$ &$0.6 \sqrt{s}$ \\
Transverse momentum &
   & & & $\B$&$<$ &$0.2$ & 
   $\pt $& $<$ &$0.1  \sqrt{s}$ & $\pt $& $<$ &$0.1  \sqrt{s}$ \\
Longitudinal momentum &
   & & & $E_{\rm lost} $& $<$& $E_1 , E_2$ &
   $\pl $&       $<$ & $E_1 , E_2$ & $\pl $&       $<$ &$E_1 , E_2$ \\
\hline
\end{tabular}
\caption{Summary of the kinematic cuts. The upper part of the table
describes the definition of the four event classes, the lower part describes 
the applied cuts. For definition of the variables see 
the text. In the case of more than three observed photons in class \dd\ 
there is no requirement on the aplanarity.}\label{cuts}\end{center}
\end{table}

All events having $\acol < 10^{\circ}$ (i.e. with the
two highest-energy clusters almost collinear) are assigned to class \ca, 
independent of the number of photon candidates.
More than 90\% of the events fall in this class.
Acollinear events ($\acol > 10^{\circ}$) are separated into three classes.
Events with $\ng = 2$ are assigned to class \cb , typically containing an
energetic photon that escapes detection near the beam axis ($|\ct |>0.97$).
Planar events ($\acop < 0.1^{\circ}$) with exactly 3 observed photon 
candidates are assigned to class \cc . The remaining events,
i.e. aplanar three-photon events 
($\acop > 0.1^{\circ}$) and events with more 
than three observed photon candidates are assigned to class \dd .
Classes \cc\ and \dd are distinguished for two reasons. 
Firstly, the main signal Monte Carlo
generator calculates only events with up to three photons, class \dd\
events are not described by this program.  
Secondly, for class \cc\ events, three-body kinematics can be used to calculate
the photon energies from the angles, to determine the
invariant masses of photon pairs. This is not possible for class \dd\ events. 
The event class definitions are summarised in Table \ref{cuts}.

\begin{samepage}
Events in all classes are required to meet the following criteria: 
\begin{itemize}
\item[\bf C1]
The energy sum of all observed ECAL clusters must be
more than 60\% of the centre-of-mass energy and the transverse momentum
of the event must be less than 10\% of the centre-of-mass energy.
\item[\bf C2]
The longitudinal momentum of the event must be smaller than the energy 
of each of the two highest-energy clusters
to ensure that the two highest-energy photons in the event are observed.
\end{itemize}
\end{samepage}

\begin{figure}[p]
   \vspace*{-1cm}
   \begin{center} \mbox{
          \epsfxsize=15.0cm
           \epsffile{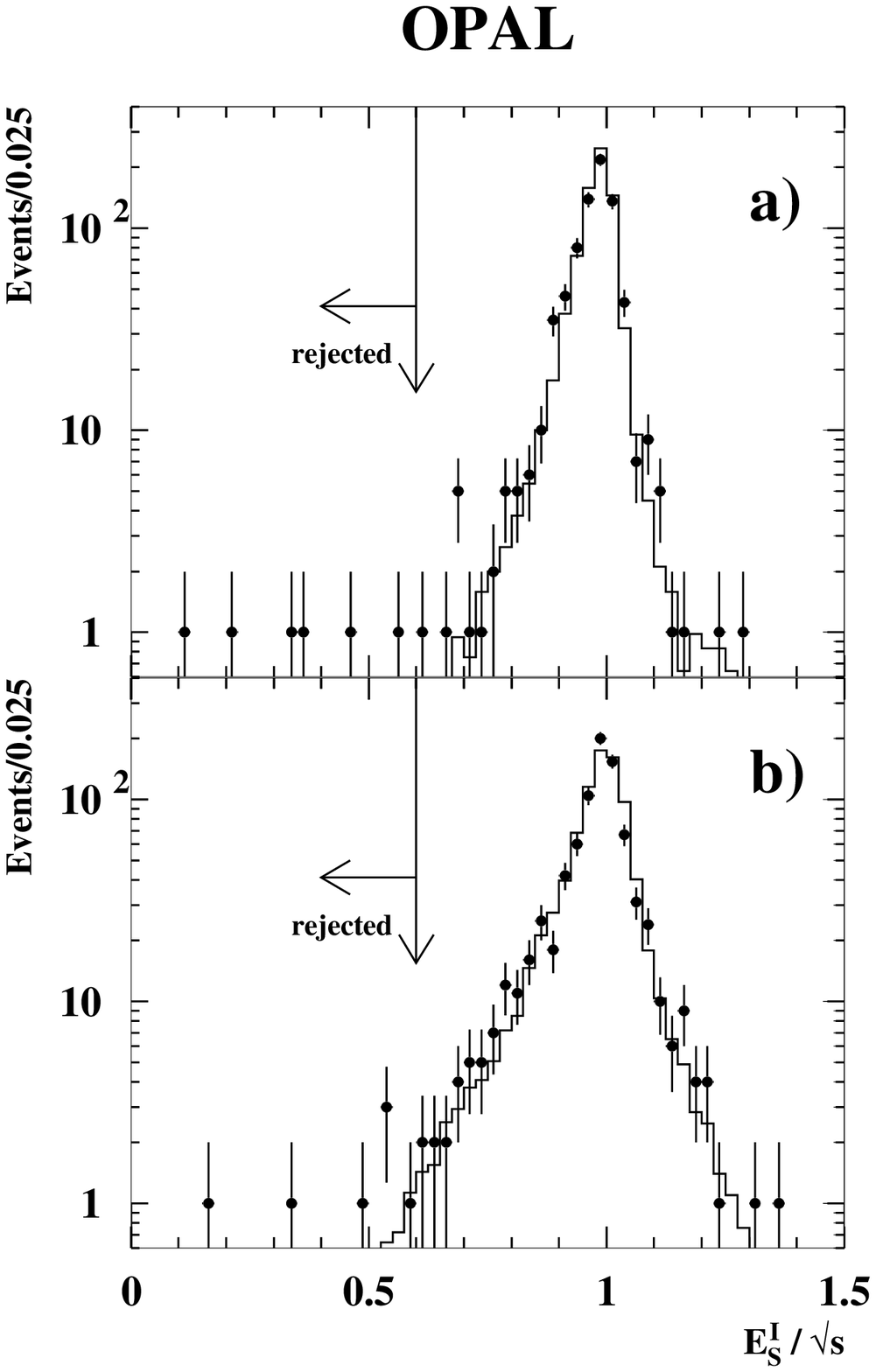}
           } \end{center}
\caption[ ]{Scaled energy sum $E_s^{I}/\sqrt{s}$ for class \ca\ events.
a) shows events with $\cte <$ 0.8 and b) shows events
with $\cte >$ 0.8.
The points show the data and the histogram represents the 
Monte Carlo expectation, normalised to the integrated luminosity 
of the data. The cut is indicated by an arrow.
}
\label{class1}
\end{figure}

For class \ca\ events it is assumed that missing energy is negligible 
due to the collinearity. There is only one cut applied to the 
sum of the two highest cluster energies $E_S^I = E_1 + E_2$. 
The distribution of $E_S^I/\sqrt{s}$ is shown in Figure \ref{class1}.
Events are selected if $E_S^I > 0.6 \sqrt{s}$.

\begin{figure}[p]
   \vspace*{-2cm}
   \begin{center} \mbox{
          \epsfxsize=15.0cm
           \epsffile{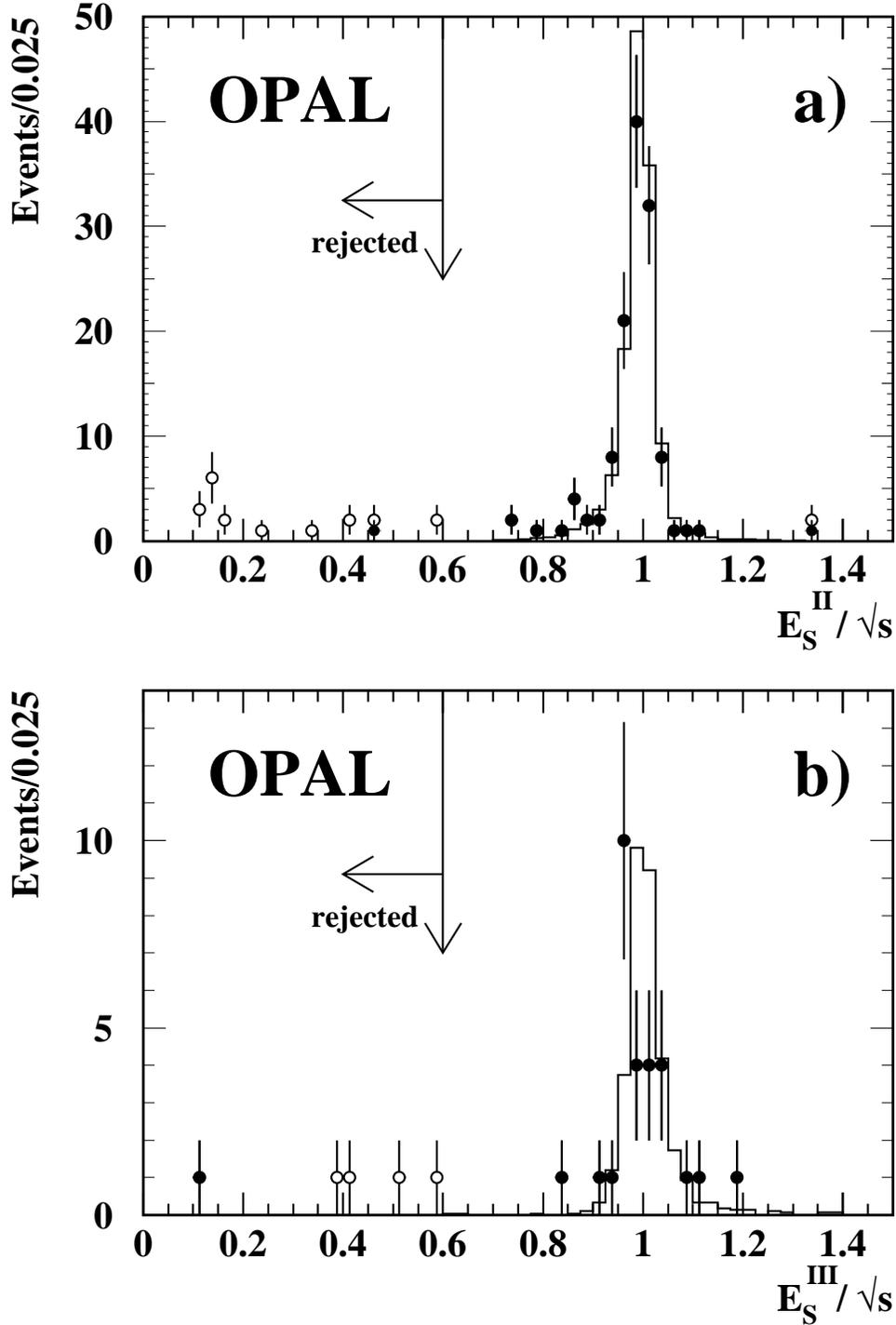}
           } \end{center}
\caption[ ]{a) Scaled energy sum $E_s^{I\!I}/\sqrt{s}$ for class \cb\ events after the
cut on $E_{\rm lost}$.
b) Scaled energy sum $E_s^{I\!I\!I}/\sqrt{s}$ for class \cc\ and \dd\
events after the cut on $\pl$.
The solid points show the data after the cut on 
the imbalance $\B$ (Fig. a) and the transverse momentum $\pt$ (Fig. b),
and the open points show the data suppressed by these requirements.
The histogram represents the Monte Carlo expectation, normalised to the
integrated luminosity of the data. The Monte Carlo distribution is shown 
after the cuts with have negligible effect. The cut is indicated by an arrow.
}
\label{class23}
\end{figure}

For class \cb\ events three-body kinematics can be used to calculate missing
energy assuming a third unobserved photon with $|\ct| =1$.
The missing longitudinal momentum can be estimated as
$E_{\rm lost} =  \sqrt{s} / \left[ 1 + (\sin{\theta_1} + \sin{\theta_2})
                       /|\sin{(\theta_1 + \theta_2)}|\right]$ 
and the transverse momentum can be approximated by the imbalance
$ \B = (\sin{\theta_1}+\sin{\theta_2}) 
\left| \cos{\left[(\phi_1-\phi_2)/2\right]}\right|$,
where $\theta_1$, $\theta_2$ and $\phi_1$, $\phi_2$
are the angles of the observed clusters. 
The energy sum $E_S^{I\!I} =  E_1 + E_2 + E_{\rm lost}$ is calculated by 
summing the two observed cluster energies and the missing energy. 
Events are selected if $E_S^{I\!I} > 0.6 \sqrt{s}$, $\B < 0.2$
and $E_{\rm lost}$ is less than both $E_1$ and $E_2$. This last requirement 
ensures that the two highest-energy photons are those observed.
Figure \ref{class23}a shows the distribution of $E_S^{I\!I}/\sqrt{s}$ 
for the data before and after the cut on $\B$, compared with the signal 
Monte Carlo.

For events in classes \cc\ and \dd , the cluster energies
must be used in addition to the cluster angles in calculating the transverse 
and longitudinal momenta ($\pt$, $\pl$) of the system. The energy sum 
$E_S^{I\!I\!I} =  \sum_{i=1}^{\ng} E_i + \pl$ is calculated by adding 
$\pl$ to the cluster energies $E_i$. 
Events are selected if $E_S^{I\!I\!I} > 0.6 \sqrt{s}$ 
and  $\pt < 0.1 \sqrt{s}$. Again, the missing energy along the beam axis, 
now determined by $\pl$, must be smaller than the energies of the two 
highest-energy clusters.
Figure \ref{class23}b shows the distribution of $E_S^{I\!I\!I}/\sqrt{s}$
for the data before and after the cut on $\pt$, compared with the signal 
Monte Carlo.

The non-physics background is reduced to a 
negligible level after the above kinematic requirements 
which are summarised in Table~\ref{cuts}.

% {\bf D) Charged event rejection:}
\subsection*{D) Charged event rejection}
Bhabha events, for example, have electromagnetic cluster 
characteristics similar to $\ggg$ events, but can be distinguished by
the presence of tracks in the central tracking chambers. 
The rejection of all events having tracks in the central tracking chambers, 
however, would lead to an efficiency loss because of photon conversions. 
Nevertheless, contributions from any channel with primary charged tracks 
must be reduced to a negligible level for this analysis.
% ^^^^^^^ unchanged ^^^^^^^^^^^^^^^^^^^^^^^^^^^^^^^^^^^^^^^^^^^^^^^^^^^
To achieve this, for each cluster, hits in the inner part of the 
drift chambers CV and CJ which are associated in $\phi$ with the cluster 
are counted.
If the number of such hits in CV (CJ) is above a $\theta$-dependent threshold,
as described in \cite{ich172}, a CV (CJ) token is assigned to the cluster.
% Two combinations of tokens can reject the event.
There are three ways to reject an event:
\begin{samepage}
\begin{enumerate}
\item
The single veto rejects events with both a CV and a CJ token assigned to 
the same cluster. 
\item
The double veto rejects events with either a CV or a CJ token assigned to
the two highest-energy clusters.
\item
Events are rejected if there is a reconstructed track
separated by more than 10$^{\circ}$ in azimuth from all photon candidates. 
\end{enumerate}
\end{samepage}

\section{Corrections and systematic errors}
\label{syserr}

\begin{table}[b]
\begin{center}
\renewcommand{\arraystretch}{1.3}
\renewcommand{\tabcolsep}{0.4em}
\begin{tabular}{|ll|cccccc|} \hline
 & Cut & Data & MC & $\ggg$ & $\rm e^+e^-(\gamma )$ & $\rm e\gamma (e)$ & 
  $\nu\overline{\nu}\g\g$  \\ \hline
A)& Preselection            & 116 129 & 90 576 & 2 297 & 82 157 & 6 078 & 44  \\
B)& Non-physics bg. reject. &  96 307 & 90 393 & 2 296 & 82 014 & 6 039 & 44  \\
C1)& Kinematic requirements &  88 715 & 87 429 & 2 273 & 79 848 & 5 305 & 3.2 \\ 
C2) &$E_{\rm lost} , \pl < E_1 , E_2$
                         &  78 782 & 79 460 & 2 109 & 77 183 &   168 & 0.05 \\
% D1) &Single veto             &   1 835 &  1 914 & 1 854 &     59 &   1.1 & 0.04 \\
% D2+3) 
D) &Charged event rejection    &   1 740 &  1 777 & 1 775 &    0.4 &   0.9 & 0.04 
\\ \hline

 &Statistical errors &           42   &    8   &   8 &    0.2 &  0.4  & 0.01 
 \\\hline
 \end{tabular}
\renewcommand{\arraystretch}{1.0}
 \caption{Number of selected events after the different cuts described in the 
text. The numbers are given for the data and the sum of Monte Carlo samples 
with the breakdown by final states given in the following columns. The BHWIDE 
generator is labelled by $\rm e^+e^-(\gamma )$ and TEEGG by $\rm e\gamma (e)$. 
All Monte Carlo predictions are normalised to the integrated luminosity of 
the data.}
 \label{eventcuts}
 \end{center}
 \end{table}

Table \ref{eventcuts} shows the number of selected events from the data,
the signal Monte Carlo and the major background sources after the different
selection cuts. Other processes contribute only a small number of events
\cite{ich183}. The preselected data have some contribution from
non-physics backgrounds and from the process $\epem\to\rm q\bar{q}$
until the kinematic cuts are applied.
There is little  efficiency loss up to this point. The restriction on the 
missing longitudinal momentum ($E_{\rm lost}$ for class \cb ) rejects events 
with a high-energy photon escaping along the beam axis. About 97\%\ of the 
remaining sample consists of Bhabha events and is well described by the 
Monte Carlo. The principal decrease in efficiency is due to the
charged event rejection.
After the final selection 1.3$\pm$0.5 background events are
expected which is negligible compared to the expected $\ggg$ signal.

The efficiency and angular resolution of the reconstruction are determined
using an $\O (\alpha^3)$ Monte Carlo sample with full detector simulation.
In the angular range $\cte < 0.8$ the efficiency is in the range 91 -- 95\%\
and drops to 40 \%\ for $\cte > 0.96$.
The requirement with the lowest efficiency is the charged event
rejection leading to a loss of events with early photon conversions.

To compare the conversion probability in data and Monte Carlo,
a special event sample is selected.
This sample consists of events with exactly 2 clusters, 
an acollinearity angle $\acol < 5^{\circ}$ and 
an energy sum $E_S^I > 0.6 \sqrt{s}$.
At least one cluster must be consistent with a non-converted photon:
neither a CV or a CJ token, nor a 
track with a minimum number of hits must be assigned to the cluster.
This leads to a sample of 2810 unbiased clusters, where the cluster
is selected via the opposite photon, i.e. all clusters, where the opposite
cluster is consistent with a non-converted photon. 
This sample is used to study the probability that a CV or CJ token is assigned 
to a cluster. There are 13.5 $\pm$ 1.5 
background events expected from Bhabha and multi-hadronic events
in the angular range $\cte < 0.8$. This may be compared to 14 events found
by scanning the data after removing an additional 8 events originating from 
radiative return to J/$\Psi$. 
Concerning the single-veto probability, 51 clusters with both a CV and a CJ 
token assigned are found in the data, in agreement with the $\ggg$ Monte Carlo 
expectation of 44 clusters. Also for
the double-veto probability for events that are not rejected by the single
veto, a good agreement between data and Monte Carlo is observed:
41 clusters with either a CV or a CJ token assigned are found in the data 
while 40 are expected from Monte Carlo in the angular range $\cte <$0.8.
In the angular range $\cte <$0.97 there are 258 clusters observed in the data
and 248 expected from Monte Carlo. The study of the data alone leads to a 
0.5\%\ systematic error per cluster on the 
single-veto probability at $\ct = 0$ due to the data statistics only.
Including the information from the Monte Carlo, the overall systematic error
on the efficiency is taken to be 1\%\ at $\ct = 0$, corresponding to 20\%\ 
of the inefficiency; this value is used for all angles.
The 1\%\ error is taken to be correlated between all $\cte$ bins.  

The veto probability due to detector occupancy is determined from randomly
triggered events. It is found to be 0.7\%\ for the charged-event rejection (D)
and 0.3\%\ for the non-physics background rejection (B).

The agreement between generated and reconstructed event angles gives
an event-angle resolution of about $0.2^{\circ}$.
In addition, the cluster angle has been compared to the
track angle for Bhabha events. For clusters with $|\ct | > 0.96$, the cluster 
angle is systematically about $0.4^{\circ}$ closer to the beam axis than the 
track angle. For clusters with $|\ct | < 0.94$, the difference is less than
$0.1^{\circ}$. Due to the cut-off at $\cte <$ 0.97 this would lead
to a decrease of the measured total cross-section by 1.1\% 
if this effect is caused by the cluster angle. 
It is included in the systematic error on the total cross-section.

The luminosity is derived from small-angle Bhabha scattering measured 
on both sides of the detector in the polar  angle region 
34 mrad $< \theta <$ 56 mrad. Uncertainties derive from the selection and the 
theoretical cross-section, as well as from a 20 MeV uncertainty on the 
beam energy. To be conservative and include additional uncertainties an 
increased systematic error of 0.5\% is taken into account for the final errors.

Since the deviations from QED (Equations \ref{lambda} -- \ref{graviton}) are 
given with respect to the Born level, the observed angular distribution is 
corrected to the Born level. The effect of radiative corrections is quantified 
by the 
ratio $\R$ of the angular distribution of the $\eeggg$ Monte Carlo and the 
Born-level cross-section:
\be \R = \xmc \!(\cte) \; \left/ \; \xb \!(\cte) \right. . \label{corrad}\ee
It is assumed that the higher-order corrections in the context of
the studied models are equal to those expected from QED. 
The ratio $\R$ varies between 1.03 and 1.06 within the studied angular range
$\cte < 0.97$ 
and is used to correct the data bin by bin to the Born level. A 1\%\ systematic 
error from ${\cal O}(\alpha^4)$ and higher-order effects is taken to be 
correlated between all bins.

\section{Results}
\subsection*{Cross-section}

\begin{table}[b]
\begin{center}
\renewcommand{\arraystretch}{1.3}
\begin{tabular}{|l|c|c|c|c|c||c||c|c|}\hline
Class & \ca & \cb & \cc & \multicolumn{2}{c||}{ \dd } & All 
 & \multicolumn{2}{c|}{Cross-section [pb]}\\
 $N_{\gamma}$ &$\geq 2$  & 2  & 3 & ~3~ & $\ge 4$ & events & $\cte < 0.9$ & $\cte < 0.97$\\
\hline
Observed & 1587 & 125 & 20 & 7  & 1   & 1740 & 
$7.55 \pm 0.18 \pm 0.14$ & $11.4 \pm 0.3 \pm 0.3$\\
Expected & 1612 & 131 & 32 & -- & 0.8 & 1776 & 7.49 & 11.8 \\ \hline
\end{tabular}
\caption[ ]{Number of observed and expected events in the angular range
$\cte <$ 0.97 for different classes. $N_{\gamma}$ is the number of observed 
photon candidates. No Monte Carlo expectation is available for class \dd\
events with 3 observed photons. The total cross-section is corrected to the 
Born level and is given for
two angular ranges to allow comparison of these results with those of
previous OPAL measurements. The first error is statistical, the second
is systematic.}
\label{events}
\end{center}
\end{table}

\begin{figure}[p]
   \begin{center} \mbox{
          \epsfxsize=17.0cm
           \epsffile{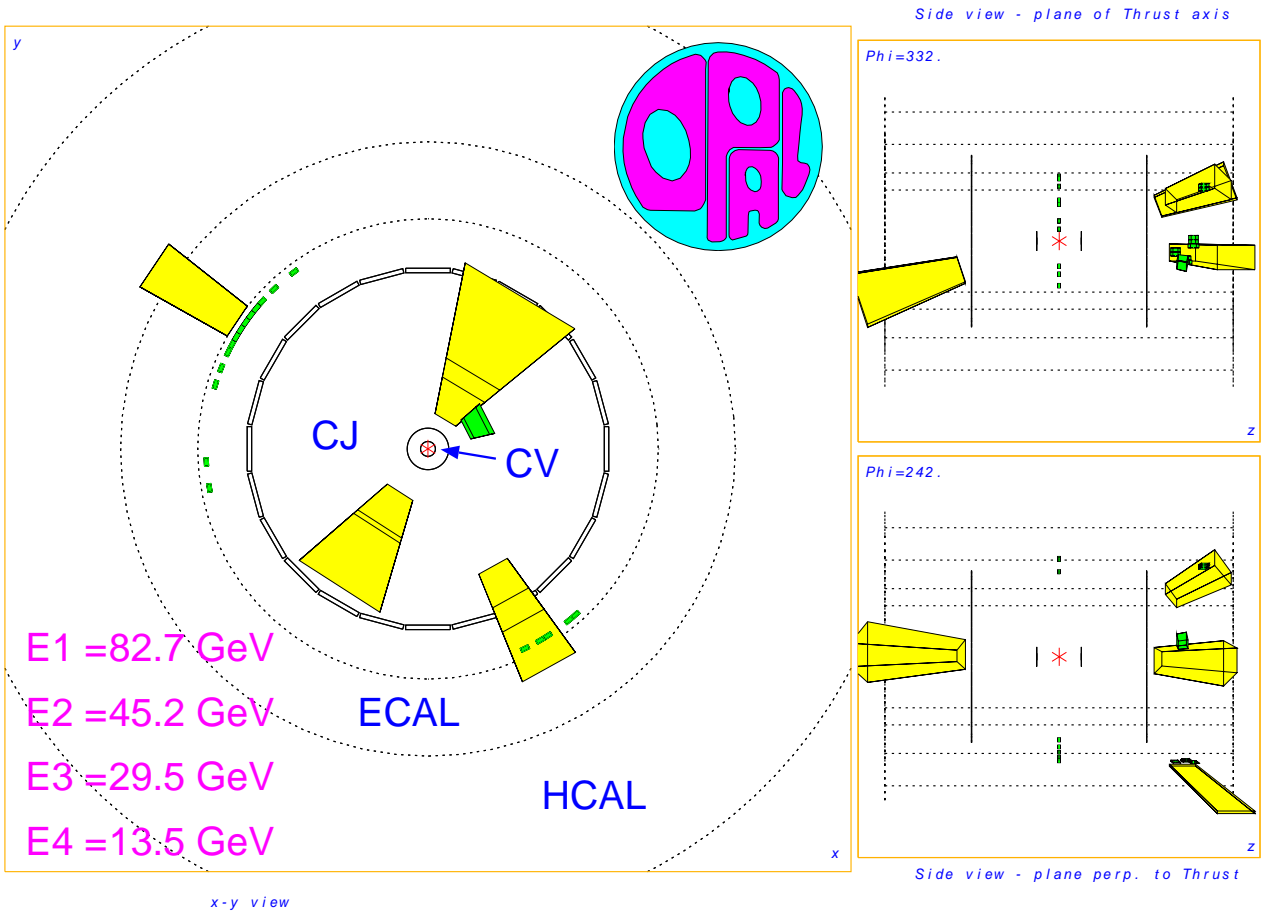}
           } \end{center}
\caption[ ]{Display of an event with four detected photons. Reconstructed
clusters in the ECAL are shown as blocks where the size is proportional
to the energy deposit. Since the two highest energy photons have an
acollinearity of 3.9$^\circ$ it is selected as class \ca\ event.
There are 0.07 events expected in class \ca\ with four  photons
with an energy of more than 10 GeV each. 
}
\label{event}
\end{figure}

\begin{figure}[p]
   \vspace*{-1cm}
   \begin{center} \mbox{
          \epsfysize=21.0cm
           \epsffile{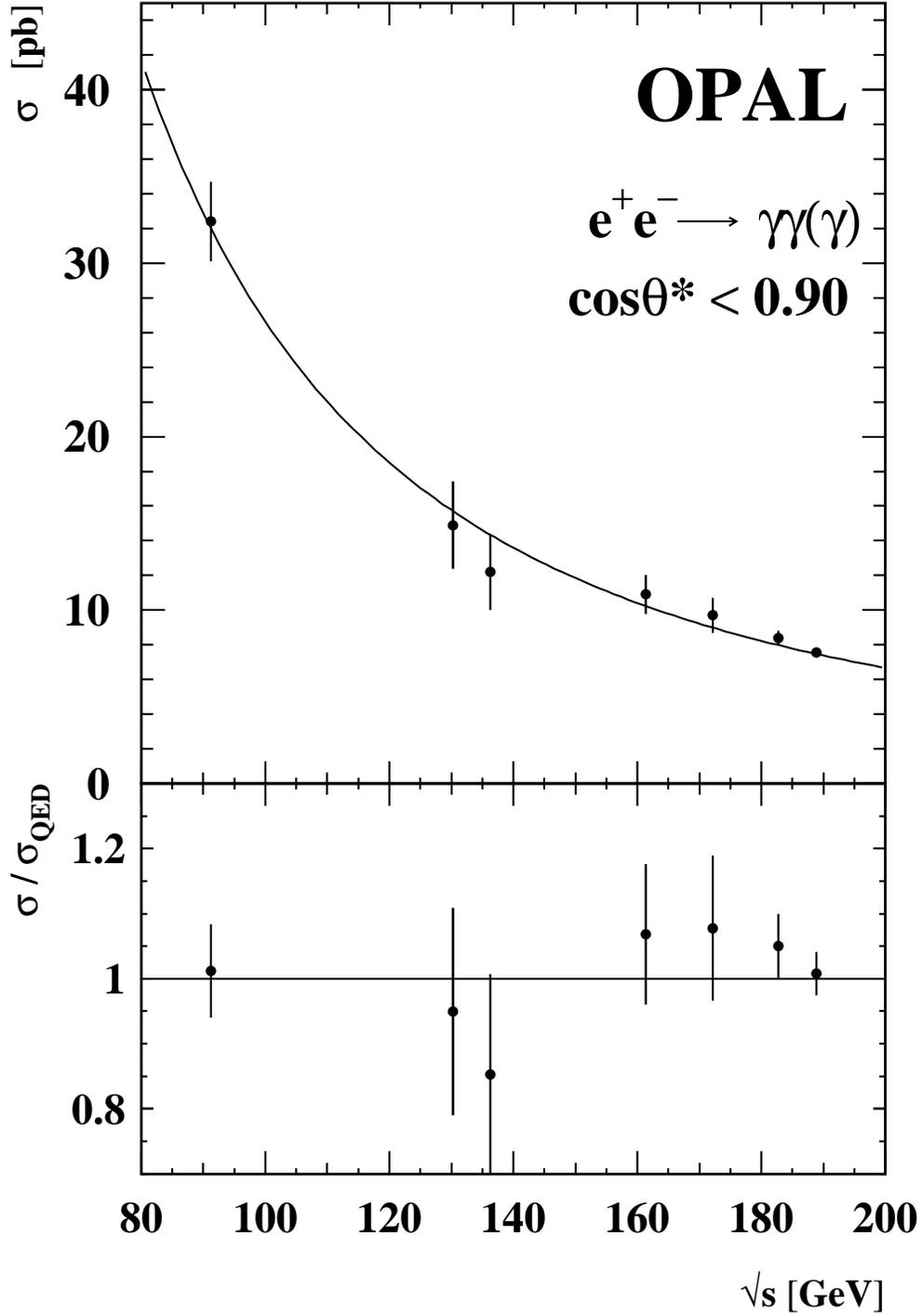}
           } \end{center}
\caption[ ]{Total cross-section for the process $\eeggg$ with $\cte < 0.9$. 
The curve corresponds to the Born-level QED expectation.
The lower plot shows the distribution normalised to the QED expectation.
The data are corrected for efficiency loss and higher-order effects and
correspond to the Born level. Results at lower energies are taken from
previous OPAL publications \cite{ich172,ich183,opalt}.
}
\label{totxsn}
\end{figure}

In Table \ref{events} the numbers of observed events in the different kinematic
classes are compared to the Monte Carlo expectation after the final selection. 
Since there is no 
Monte Carlo generator for the case of four-photon events where at least one 
photon is along the beam axis, no prediction for class \dd\ events with three 
observed photons is given. The prediction for events with
four or more observed photons is calculated using FGAM. 
One class \dd\ event with 4 detected photons is observed. Figure \ref{event}
shows a class \ca\ event with 4 detected photons, where the lowest-energy 
photon has an energy of 13.5 GeV. The expectation for 
a class \ca\ event with four photons with at least 10 GeV each is 0.07 events.
The total cross-section $\sigma$ for the process $\eeggg$, determined
from 1740 events selected in the range $\cte < 0.97$, is also given 
in Table \ref{events}. 
The cross-section is corrected for detection efficiency, $\O (\alpha^3)$
effects derived from $\R$ (Eq. \ref{corrad}) and the veto probability
due to detector occupancy.
The cross-section at the different LEP energies as measured by OPAL in 
the range $\cte < 0.9$ is shown in Figure \ref{totxsn}. All measurements 
are in good agreement with the QED prediction.

\begin{figure}[p]
   \vspace*{-1cm}
   \begin{center}
      \mbox{
          \epsfxsize=16.0cm
          \epsffile{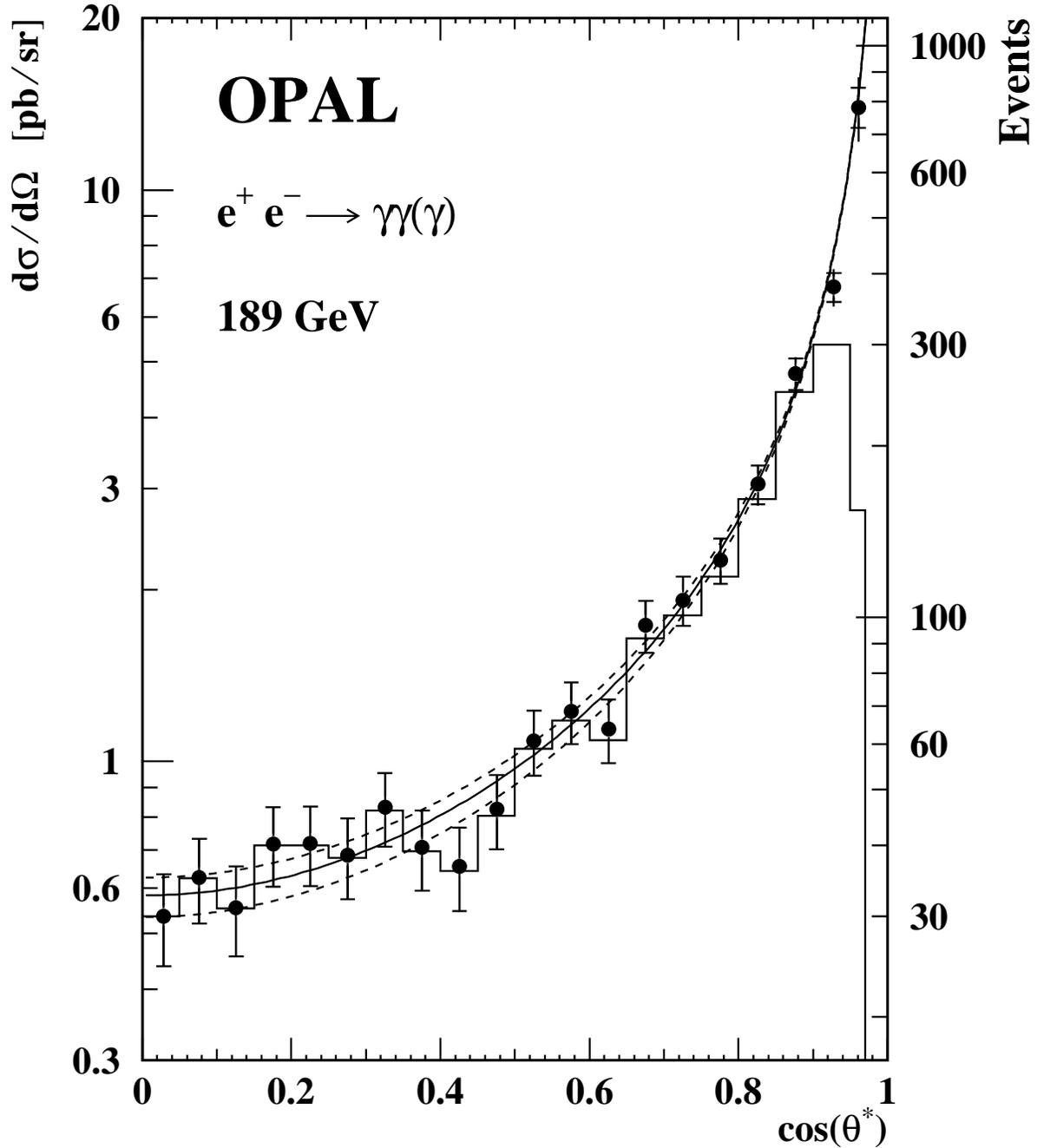}
           }
   \end{center}
\caption[ ]{The measured angular distribution for the process 
$\epem\to\g\g(\g)$ at $\sqrt{s} = $ 189 GeV. The histogram shows the observed 
number of events per bin. Note the smaller width of the highest $\cte$ bin.
The points show the number of events corrected for efficiency and radiative
effects. The inner error-bars correspond to the statistical error and the 
outer error-bars to the total error. The solid curve corresponds to
the Born-level QED prediction. The dashed lines represent the 95\% CL
interval of the fit to the function $\xl$.  }
\label{wq}
\end{figure}

The angular distribution of the observed events and the measured differential 
cross-section obtained by applying efficiency and radiative corrections, are 
shown in Figure \ref{wq}. The data have a $\rm\chi^2 /NDF = 14.8/20$ with 
respect to the QED expectation (solid line). The 95\%\ CL interval of a 
$\chi^2$-fit to the data of the function $\xl$ (Eq. \ref{lambda}) is also 
shown (dashed line). To obtain the limits at 95\% confidence level, the
probability is normalised to the physically allowed region,
i.e. $\Lambda_+ > 0$ and $\Lambda_- < 0$ as described in Ref. \cite{pdg}.
For both functions $\xl$ and $\xq$ the $\chi^2$ distribution is parabolic
as a function of the chosen fitting parameters $\Lambda_\pm^{-4}$ and 
$\Lambda'^{-6}$. 
The $\chi^2$ distribution for $\xg$ from graviton exchange in the context
of extra dimensions is also approximately parabolic.
The asymmetric limits $x_{95}^{\pm}$ 
on the fitting parameter are obtained by:
\be \frac{\int^{x_{95}^+}_0 \Gamma(x,\mu ,\sigma ) dx}
         {\int^{\infty   }_0 \Gamma(x,\mu ,\sigma ) dx} = 0.95 \; 
    \hspace{7mm}\mbox{and}\hspace{7mm}
    \frac{\int_{x_{95}^-}^0 \Gamma(x,\mu ,\sigma ) dx}
         {\int_{-\infty  }^0 \Gamma(x,\mu ,\sigma ) dx} = 0.95 \; , 
         \label{limeq}\ee
where $\Gamma$ is a Gaussian with the central value and error of the fit
result denoted by $\mu$ and $\sigma$, respectively. This is equivalent
to the integration of a Gaussian probability function as a
function of the fit parameter. It is not possible to integrate the 
probability function as a function of $\Lambda_\pm$, $\Lambda '$ or $G_\pm$.
Therefore the limits on the fit parameters are derived via Eq. \ref{limeq}.
The 95 \% CL limits on the cut-off parameters are derived from the limits on the 
fit parameters, e.g. the limit on $G_+$ is obtained as 
$[x_{95}^+(G^{-4}_{\pm})]^{-1/4}$.

\begin{figure}[p]
   \begin{center}
      \mbox{
          \epsfxsize=17.0cm
          \epsffile{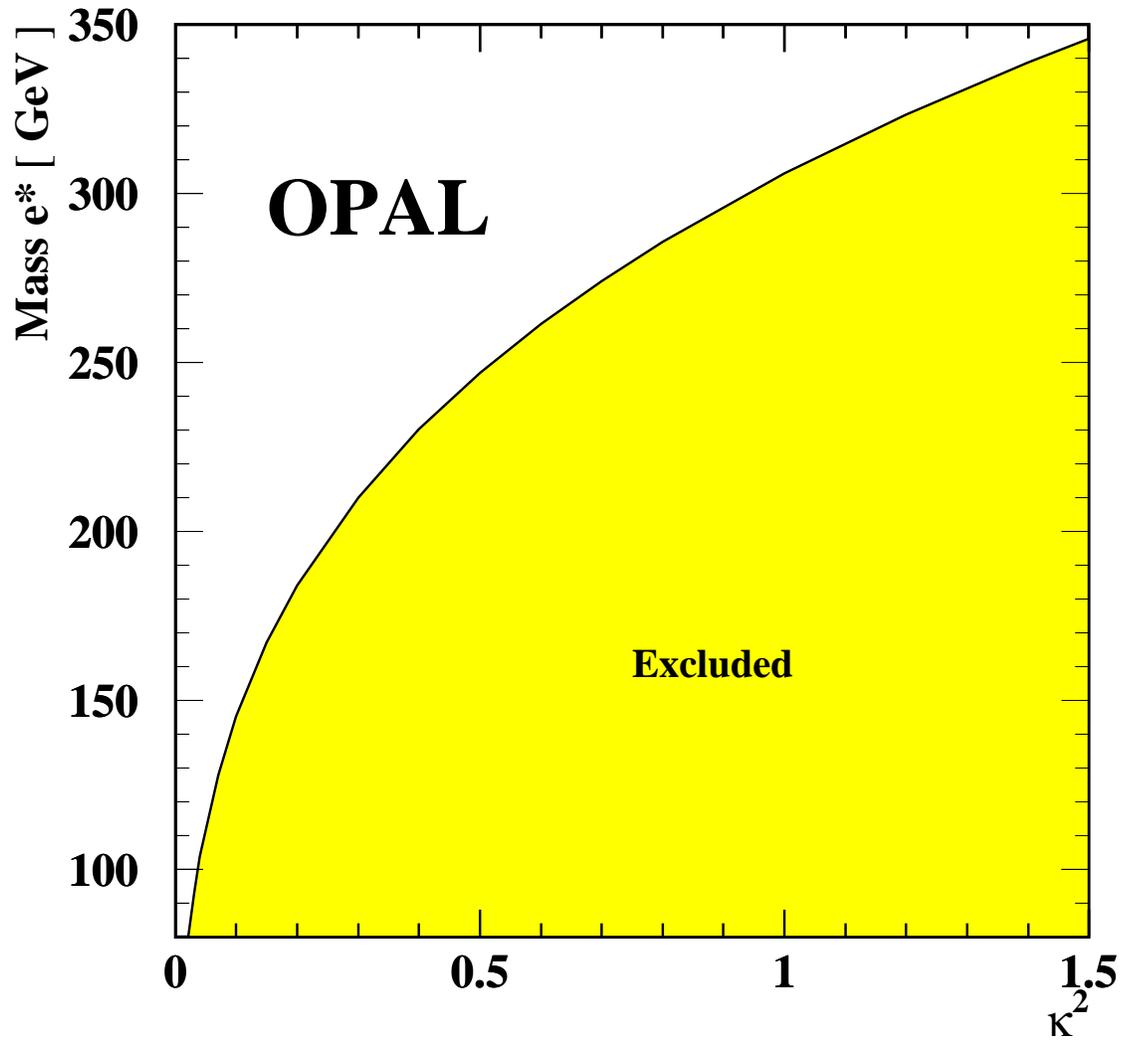}
           }
   \end{center}
\caption[ ]{Lower limit (95\%\ CL) on 
the mass $M_{\rm e^{\ast}}$ of an excited electron as a function of
the square of the $\rm e^{\ast}e\gamma$ coupling constant ratio 
$\kappa^2$ obtained from a combined fit to data taken at 183 and 189 GeV.}
\label{elimit}
\end{figure}

The limit on the mass of an excited electron $M_{\rm e^{\ast}}$ as a function
of the coupling-constant ratio $\kappa$ for the ($\mathrm{e^*e\g}$)-vertex, 
which is fixed during the fit, is shown in Figure \ref{elimit}. In the case of 
$\xe$ the cross-section does not depend linearly on the chosen fitting 
parameter $M_{\rm e^{\ast}}^{-2}$ and the limit corresponds to
an increase of the $\chi^2$ by 3.84 with respect to the minimum.

\begin{table}[b]
\begin{center}
\renewcommand{\arraystretch}{1.5}
\begin{tabular}{|c|c|ccc|}\hline
& & &\multicolumn{2}{c|}{ 95\%\ CL Limit [GeV] }\\ 
Fit parameter & Fit result (189 GeV)& & 189 GeV & 183+189 GeV \\ \hline
 &  & $\Lambda_+ >$ & 297  & 304 \\
 \raisebox{2.2ex}[-2.2ex]{$\Lpm^{-4}$} &
 \raisebox{2.2ex}[-2.2ex]{$(-36 \pm 71)\cdot 10^{-12}$ GeV$^{-4}$}
 & $\Lambda_- > $&  281 & 295   \\
$\Lambda '^{-6}$ & $ (-2.7 \pm 6.3 )\cdot
                                  10^{-18}$ GeV$^{-6}$
& $\Lambda '~ > $&  674 & 672   \\
 &  & $G_+ >$ &  660  & 660 \\
 \raisebox{2.2ex}[-2.2ex]{$G^{-4}_{\pm}$} &
 \raisebox{2.2ex}[-2.2ex]{ $ \left(-1.7{+3.2 \atop -3.3}\right)\cdot
                                         10^{-12}$ GeV$^{-4} $ }
 & $G_- > $&  600 & 634    \\
%  $M_{\rm e^{\ast}}^{-2}$ & $ \left(0.0{+6.6 \atop -19.4}\right)\cdot
$M_{\rm e^{\ast}}^{-2}$ & $ (0.0 \pm 6.6)\cdot
                                         10^{-6}$ GeV$^{-2} $
& $M_{\rm e^{\ast}} > $&   311 & 306  \\ \hline
\end{tabular}
\renewcommand{\arraystretch}{1.0}
\caption[ ]{Fit results and 95\% CL lower limits obtained from the fit to the
differential cross-section. The last column shows the limit obtained from
a combined fit to the data taken at 183 and 189 GeV.
The limit for the mass of an excited electron is determined
assuming the coupling-constant ratio $\kappa = 1$. }
\label{result}
\end{center}
\end{table}

The fit results are summarised in Table \ref{result}. Limits obtained from a
combined fit to the data taken at 189 and 183 GeV are also given.
The limits obtained are 30 -- 115 GeV higher than our previous results
\cite{ich183}. 
% Similar limits are obtained by ALEPH and DELPHI at 183 GeV \cite{aleph183}.

\subsection*{Resonance production}

\begin{figure}[p]
   \vspace*{-2cm}
   \begin{center} \mbox{
          \epsfxsize=14.0cm
           \epsffile{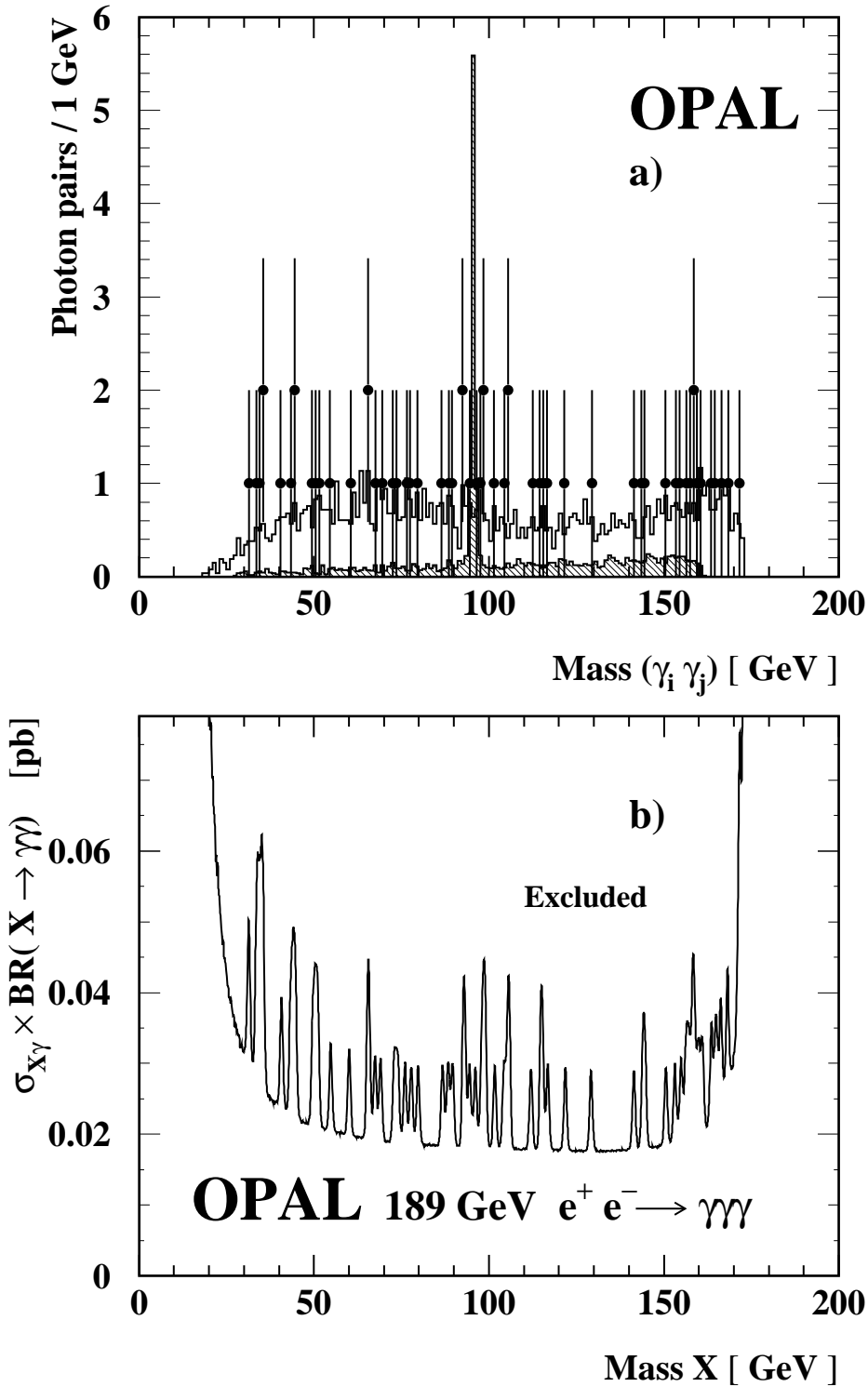}
           } \end{center}
\caption[ ]{Results of a search for resonance production in class 
\cc\ events.
a) shows the invariant mass of photon pairs for data (points) 
and for the $\eeggg$ Monte Carlo expectation (open histogram).
The hatched histogram represents a $\gamma\gamma$ resonance at 95.5~GeV
with a cross-section times branching ratio of 0.05 pb. 
The 
binning is chosen to match the expected mass resolution.
b) shows the upper limit (95\%\ CL) for the cross-section times 
branching ratio for the process $\rm \epem\to X \g$, $\rm X\to\g\g$ 
as a function of the mass of the resonance X. }
\label{masslim}
\end{figure}

A resonance X produced by the process $\epem\to{\rm X}\g$ and decaying into 
two photons, ${\rm X} \to \g\g$, would be visible in the two-photon invariant mass
spectrum, since this process leads to a three-photon final state without missing
energy. Searches for such a resonance have been performed previously at the 
Z$^0$ peak \cite{gres} and at higher energies \cite{ich183,l3,delhiggs}, 
leading to bounds on Higgs and gauge boson interactions \cite{eboli2}.
For this search, 20 events from class \cc\ are used.
The invariant mass of each photon pair is shown in Figure \ref{masslim}a).
The energies of the three photons are not based on the measured 
cluster energies but are calculated from the photon angles assuming
three-body kinematics:
\be
E_k \propto \sin{\alpha_{ij}} \; ; \;  E_1 + E_2 + E_3 = \sqrt{s} ,
\ee
with $E_k$ the energy of one photon and $\alpha_{ij}$ the angle between the
other two photons. In this case, the typical mass resolution for photon pairs is
about 0.5 GeV. Deriving the mass from the measured cluster energies would lead
to a resolution of about 3\%.
The distribution is consistent with the 
Monte Carlo expectation from the QED process $\eeggg$. There is no 
evidence for an enhancement due to a resonance.

An upper limit on the total production cross-section multiplied by the
two-photon branching ratio is calculated using the method of Ref. \cite{bock}. 
This method uses fractional event counting where the weights assigned to each
photon pair depend on the expected resolution and the difference between the 
hypothetical and the reconstructed mass. The limits shown in 
Figure \ref{masslim}b) are obtained assuming the 
natural width of the resonance to be negligible. The $\eeggg$ background is 
subtracted. For the efficiency correction, the production and subsequent decay 
of the resonance are assumed to be isotropic.
The mass range is limited by the acollinearity restriction.
Regarding a model with anomalous couplings of the Higgs boson \cite{eboli2},
this analysis gives access to a larger mass range than the analysis of 
$\epem\to\rm H Z$ with $\rm H \to \gamma\gamma$ \cite{opalhiggs}.
However, in the region of overlap the direct search is more sensitive.

\section{Conclusions}

The process $\eeggg$ has been studied using data taken with the OPAL
detector at centre-of-mass energies up to 189 GeV.
The measured angular distribution and total cross-section for
this process both agree well with QED predictions. The
limits (95\%\ CL) on cut-off parameters are $\Lambda_+ > $ 304 GeV,
$\Lambda_- > $ 295 GeV and  $\Lambda ' > $ 672 GeV. An excited
electron is excluded for $M_{\rm e^{\ast}} <$ 306 GeV assuming the 
$\rm e^{\ast}e\gamma$ and $\rm ee\gamma$ coupling to be the same. 
Graviton exchange in the context of theories including extra dimensions 
is excluded for scales $G_+ <$~660~GeV and $G_- <$~634~GeV. In the 
$\gamma\gamma$ invariant mass spectrum of events with three 
final-state photons, no evidence is found for a resonance X decaying to $\g\g$.
A limit on the production cross-section times branching ratio is derived as a 
function of the mass $M_{\rm X}$. 

\section*{Acknowledgements}

We particularly wish to thank the SL Division for the efficient operation
of the LEP accelerator at all energies
 and for their continuing close cooperation with
our experimental group.  We thank our colleagues from CEA, DAPNIA/SPP,
CE-Saclay for their efforts over the years on the time-of-flight and trigger
systems which we continue to use.  In addition to the support staff at our own
institutions we are pleased to acknowledge the  \\
Department of Energy, USA, \\
National Science Foundation, USA, \\
Particle Physics and Astronomy Research Council, UK, \\
Natural Sciences and Engineering Research Council, Canada, \\
Israel Science Foundation, administered by the Israel
Academy of Science and Humanities, \\
Minerva Gesellschaft, \\
Benoziyo Center for High Energy Physics,\\
Japanese Ministry of Education, Science and Culture (the
Monbusho) and a grant under the Monbusho International
Science Research Program,\\
Japanese Society for the Promotion of Science (JSPS),\\
German Israeli Bi-national Science Foundation (GIF), \\
Bundesministerium f\"ur Bildung, Wissenschaft,
Forschung und Technologie, Germany, \\
National Research Council of Canada, \\
Research Corporation, USA,\\
Hungarian Foundation for Scientific Research, OTKA T-029328, 
T023793 and OTKA F-023259.\\

\clearpage

\end{document}